\documentclass[aip,twocolumn,floatfix,10pt]{revtex4-2}

\usepackage[utf8]{inputenc}
\usepackage[english]{babel}
\usepackage{graphicx}
\usepackage{amsmath,amssymb}
\usepackage{amsfonts}
\usepackage{mathrsfs}
\usepackage{physics}
\usepackage[colorlinks=true,hyperfootnotes=true,breaklinks=true,citecolor=red,urlcolor=blue,linkcolor=blue]{hyperref}
\usepackage{mathbbol} 
\usepackage{forest}
\usepackage{tikz}
\usetikzlibrary{positioning}
\usepackage{soul}
\begin{document}
\title{Optimizing topology for quantum probing with discrete-time quantum walks}
\author{Simone Cavazzoni}
\email{simone.cavazzoni@unimore.it}
\affiliation{Dipartimento di Scienze Fisiche, Informatiche e Matematiche,  Universit\`{a} di Modena e Reggio Emilia, I-41125 Modena, Italy}
\author{Paolo Bordone}
\email{paolo.bordone@unimore.it}
\affiliation{Dipartimento di Scienze Fisiche, Informatiche e Matematiche, Universit\`{a} di Modena e Reggio Emilia, I-41125 Modena, Italy}
\affiliation{Centro S3, CNR-Istituto di Nanoscienze, I-41125 Modena, Italy}
\author{Matteo G. A. Paris}
\email{matteo.paris@fisica.unimi.it}
\affiliation{Quantum Technology Lab, Universit\`{a} di Milano, I-20133 Milano, Italy}
\date{\today}
\begin{abstract}
Discrete-time quantum walk (DTQW) represents a convenient mathematical framework for describing the motion of a particle on a discrete set of positions when this motion is conditioned by the values of certain internal degrees of freedom, which are usually referred to as the {\em coin} of the particle. 
As such, and owing to the inherent dependence of the position distribution on the coin degrees of freedom, DTQWs naturally emerge as promising candidates for quantum metrology.
In this paper, we explore the use of DTQWs as quantum probes in scenarios where the parameter of interest is encoded in the internal degree of freedom of the walker, and investigate the role of the topology of the walker's space on the attainable precision. 
In particular, we start considering the encoding of the parameter by rotations for a walker on the line, and evaluate the quantum Fisher information (QFI) and the position Fisher information (FI), explicitly determining the optimal initial state in position space that maximizes the QFI across all encoding schemes. 
This allows us to understand the role of interference in the position space and
to introduce an optimal topology, which maximizes the QFI of the coin parameter and makes the position FI equal to the QFI. 
\end{abstract}
\maketitle

\section{Introduction}
\label{sec:Introduction}
Discrete-Time Quantum Walks provide a versatile physical model to describe a large variety of physical systems and phenomena \cite{childs2009universal,lovett2010universal,xie2020topological,kadian2021quantum,apers2022quadratic}. Defined in a discrete spacetime, they are the quantum analogue of classical random walks \cite{aharonov2001quantum,wong2017coined}. To completely characterize the dynamics of the system it is essential to know the initial state of the walker 
in the discrete position space and in its internal degree of freedom, usually referred to as the coin space. The time evolution is performed by a unitary operator $\mathcal{U}$, which evolves the state of the walker accordingly to the coin state and creates entanglement between the spatial and internal degree of freedom of the walker. This operator is defined through the {coin operator} $\mathcal{C}$ that acts on the internal degree of freedom of the walker and through the {conditional shift operator} $\mathcal{S}$, which modify the position of the walker according to the coin state. 

Experimental implementations of DTQWs have been realized in photonic systems\cite{zhu2020photonic,di2023ultra}, spin systems \cite{witthaut2010quantum} and in trapped atoms \cite{karski2009quantum}. Recently the DTQW paradigm has also been implemented in quantum computers \cite{acasiete2020implementation,wing2023circuit}. Moreover {DTQW} have found appliance also in modern technological developments, such as state transfer and quantum routing \cite{li2021discrete} or in communication protocols \cite{srikara2020quantum}. Due the flexibility of this model, {DTQW} have been studied as a resource of universal quantum computing \cite{singh2021universal,Chawla2023} and in quantum algorithm applications \cite{shakeel2020efficient}. Morever, the dependence of the position of the walker on its internal degrees of freedom, makes any measurement of the position of the particle conditioned by both the coin state and operator. Accordingly, {DTQWs} scheme found applications in metrological studies \cite{singh2019quantum,annabestani2022multiparameter,PhysRevA.92.010302}. Other works have found a connection between {DTQW} and measurement procedures \cite{wang2023generalized,tornow2023measurement}, proving the intrinsic and profound relation between {DTQW} and metrology. Using the walker as a probe and measuring the position of the {walker} it is possible to extract information about the parameters encoded in the coin's degrees of freedom.

In this work, we study in details the role of interference on the 
time evolution of the system and use this information to individuate the key features to maximize the quantities of metrological interest, 
such as the quantum and classical Fisher information. Then, we exploit this information to design the optimal graph topology, suitable for estimation protocols. In systems where the topology of the position space can be 
controlled, such as optical wave-guides 
\cite{neves2018photonic,zhang2018quantum}, and quantum circuits \cite{douglas2009efficient,acasiete2020implementation,nzongani2023quantum,wing2023circuit}, our scheme may be conveniently adopted.  Our work is also motivated by the rising interest in experimental realization of 
QWs with various topologies in systems and materials developed for modern electronics \cite{qiang2021implementing} and quantum technologies \cite{dimcovic2011framework,asaka2021quantum,asaka2023two1,asaka2023two2}. 

The paper is organized as follow. In Sec. \ref{sec:DTQW} we introduce a 
suitable  mathematical framework for {DTQW}, including the time evolution operator and the initial state of the walker. Then, in Sec. \ref{sec:Metrology} we briefly review the basic principles of classical and quantum metrology, introducing the Fisher information for position measurements and the quantum Fisher information. In Sec. \ref{sec:Interference} and Sec.\ref{sec:Metrology} we present and discuss our results. We starting studying the role of interference 
on the (Q)FI and then maximize the QFI of the system for different coin encodings. Finally, we design the ideal topology for estimation problems involving {DTQW}. In Sec. \ref{sec:Summary} our key results are summarized and adapted to a number of potential applications. In Appendices \ref{app:Topology}-\ref{app:Basis_Change} mathematical details about the position and coin operator that define the time evolution operator are provided. Finally, a discussion on the relationship between our results and those obtainable in qubit systems is reported in Appendix \ref{app:relation_DTQW_QuDit}.

\section{Discrete-Time Quantum Walks}
\label{sec:DTQW}

The {DTQW} paradigm defines the time evolution of the {walker}, according to his internal degrees of freedom (i.e. {coin}) over a discrete space. Both the position and coin space have an associated Hilbert space $\mathscr{H}_{p}$ and $\mathscr{H}_{c}$, with $\mathscr{H}_{p} = \operatorname{span}\{ \vert x \rangle_{p} \}$ and $\mathscr{H}_{c} = \operatorname{span}\{ \vert m \rangle_{c} \}$. The states $\vert x \rangle_{p}$ are the lattice site of a $\mathcal{N}$ dimensional discrete space (i.e. a graph $\mathcal{G}$), while $\vert m \rangle_{c}$ are usually the eigenvector of the $D$-dimensional spin operator $S_{z}$. Globally the Hilbert space of the whole system ({walker}+coin) is then $\mathscr{H} = \mathscr{H}_{p} \otimes \mathscr{H}_{c}$.

\subsection{The time evolution operator}
The single step evolution of the most general {DTQW} is described by the unitary operator, usually referred to as the conditional shift operator, 
\begin{equation}
    \label{time_evolution_operator}
    \mathcal{U} = \mathcal{S} \left( \mathbb{1}_{p} \otimes \mathcal{C}  \right),
\end{equation}
where $\mathbb{1}_{p}$ define the identity in position space and $\mathcal{C}$ is the coin operator, which acts exclusively on the internal degree of freedom of the walker.  
The most general operator $\mathcal{C}$ for a $D$ dimensional internal degree of freedom is a $U(D)$ matrix and depends on $D^2$ different parameters. The condition on the coin matrix ensure the normalization of the wave-function of the system during the time evolution of the walker. The conditional shift operator may be written as 
\begin{equation}
    \label{eq:conditional_shift_operator}
    \mathcal{S} = \sum_{x} \sum_{m} \vert f\left( x,t,m \right) \rangle_{p} \;_{p}\langle x \vert \otimes \vert m \rangle_{c} \;_{c}\langle m \vert,
\end{equation}
where $\{ \vert m \rangle_{c} \}$ are  the eigenstates of the spin operator $S_{z}$ (see \ref{app:Topology} for an additional analysis of the elements that define the conditional shift operator). The projectors $ \vert f\left( x,t,m \right) \rangle \langle x \vert $ intrinsically define the topology of the discrete space in which the walker evolves in time. Concerning the indices of the coin space, 
we use the notation

\begin{equation}
    \mathcal{M}_c^{(s)} = {m_{c}} =
    \begin{cases}
     \{ -s,...,-1,0,1,..., s \} & \text{(even $D$)},\\
	\{ -s-\frac{1}{2},...,-1,1,..., s+\frac{1}{2} \} & \text{(odd D)},
    \end{cases}
    \label{eq:set_indices_spin_coin}
\end{equation}
with $s=(D-1)/2$. Under the assumption of a unique and time independent coin operator the time evolution of the wave-function of the walker is given by
\begin{equation}
    \label{eq:time_evolution_psi}
    \vert \psi(t) \rangle = \mathcal{U} \vert \psi \left( t-1 \right) \rangle  = \mathcal{U}^{t} \vert \psi ( 0 ) \rangle,
\end{equation}
where $\vert \psi (0) \rangle$ denotes the initial state of the system. 

In this work, we consider the problem of estimating an unknown parameter $\theta$, 
encoded in the coin, as $\mathcal{C}\left( \theta \right)$. In particular, we consider the shift parameter describing a coin rotation around a fixed axis, and focus on the rotations around $x$, $y$, and $z$ axes (see \ref{app:Basis_Change} for additional information about the role of the operator 
$\mathcal{C}$ on the walker's time evolution). Our main goal is to optimize the estimation 
of $\theta$ examining the impact of both the interference and the topology of the position space on the estimation problem.

\subsection{The state of the walker}
The most general pure initial state of a walker can be written as

\begin{equation}
    \label{eq:initial_state}
    \vert \psi \left( 0 \right) \rangle = \sum_{x \in \mathcal{G}} \vert x \rangle_{p} \otimes \vert \phi_{x} \rangle_{c},
\end{equation}
where $\vert \phi_{x} \rangle_{c}$ is the coin state associated to every lattice site of the graph $\mathcal{G}$ (i.e. $\vert x \rangle_{p}$). Due to the bi-partition of the system the global wave-function has two main degrees of freedom: the distribution in position space and, for graph sites, the coin wave function. The sum over the position space ensure to include all the possible states, from the perfectly localized initial probe, in which only one site $\vert x_{0} \rangle$ is occupied, to the completely delocalized state, in which every reticular site has an associated wave-function component different from zero. The overall normalization condition reads

\begin{equation}
    \label{eq:normalization_condition}
    \sum_{x \in \mathcal{G}} \;_{c}\langle \phi_{x} \vert \phi_{x} \rangle_{c} = 1.
\end{equation}
The relations among the $\vert \phi_{x} \rangle_{c}$ define the degree of entanglement of the state. According to the dimensionality $D$ of the coin space, each coefficient $\vert \phi_{x} \rangle_{c}$ can have $D$ components. Then, at each reticular site there are $2(D-1)$ degrees of freedom that define a {qudit} state. The independent degrees of freedom are $2(D-1)$ because the normalization condition and the independence of the physics from an overall phase reduces by a factor $2$ the number of independent components $2D$ of the complex wave-function. Different parameterizations have been proposed during the years for such states. The most known is the Bloch sphere that describe a bi-dimensional space through 2 angles, defining a quantum pure state as a point on a spherical surface.

\section{Walker's Metrology}
\label{sec:Metrology}
In a {DTQW} the position of the {walker} is conditioned by the state and evolution of the coin, and thus a position measurement provides 
information about the coin parameter $\theta$ \cite{singh2019quantum,annabestani2022multiparameter}. From a classical point of view the amount of information that a position measurement $P$ carries about the unknown parameter $\theta$ is expressed through the Fisher Information as 
\begin{equation}
    \label{eq:fisher_information}
    F_{p}(\theta) = \sum_{x \in \mathcal{G}} \frac{\left[ \partial_{\theta} p( x \vert \theta ) \right]^{2}}{p\left(x \vert \theta \right)},
\end{equation}
where $\mathcal{G}$ denotes the discrete position space and $p\left( x \vert \theta \right)$ is the position probability distribution of the walker, as obtained by ignoring the coin degrees of freedom:
\begin{align}
\label{eq:p}
     p\left( x \vert \theta \right) & = 
     \hbox{Tr}\Big[ |x\rangle\langle x| \otimes \mathbb{1}_{p} \,  \vert \psi(t) \rangle \langle \psi(t) \vert \Big] \notag \\ & = 
     \langle x \vert \Tr_{c}\Big[\vert \psi(t) \rangle \langle \psi(t) \vert\Big] \vert x \rangle\,.
\end{align}

The ultimate precision attainable after $M$ measurements, and using a suitable estimator, is given by the Cram\'er-Rao bound 
\begin{equation}
    \label{eq:Cramer_Rao}
    \operatorname{Var}\left( \theta \right) \geq \dfrac{1}{M F_{p}\left( \theta \right)}.
\end{equation}
In the quantum realm the quantity
\begin{equation}
    \label{eq:QFI_def_geq_FI}
    H(\theta) = \Tr[\rho_{\theta}\mathcal{L}_{\theta}^2]
\end{equation}
defines the quantum counterpart of the Fisher information, where $\mathcal{L}$ is the symmetric logarithmic derivative (SLD), related to the derivative of the density matrix of the quantum state through the implicit relation \cite{paris2009quantum}

\begin{equation}
    \label{eq:sld}
    \partial_{\theta} \rho_{\theta} = \frac{1}{2} \{ \mathcal{L}_{\theta},\rho_{\theta} \},
\end{equation}
with $\{,\}$ denoting the anti-commutator. For pure states the SLD is given by
$\mathcal{L}_{\theta} = 2 \left( \vert \psi_{\theta} \rangle \langle \partial_{\theta} \psi_{\theta} \vert  + \vert \partial_{\theta} \psi_{\theta} \rangle\langle  \psi_{\theta} \vert \right)$ and the QFI $H(\theta)$ may be written in terms of the wave-function and its derivative as follows
\begin{equation}
    \label{eq:quantum_fisher_information_pure_states}
    H(\theta) = 4 \left( \langle \partial_{\theta} \psi_{\theta} \vert \partial_{\theta} \psi_{\theta} \rangle - \left| \langle \partial_{\theta} \psi_{\theta} \vert \psi_{\theta} \rangle\right|^2 \right),
\end{equation}
where the term $\langle \partial_{\theta} \psi_{\theta} \vert \psi_{\theta} \rangle$ is purely imaginary. The QFI is the supremum of the Fisher information, i.e. the maximum of the FI over all the possible measurement. Consequently, the quantum Cramer Rao relation is
\begin{equation}
    \label{eq:Quantum_Cramer_Rao}
    \operatorname{Var}\left( \theta \right) \geq \dfrac{1}{M F_{p}\left( \theta \right)} \geq \dfrac{1}{M H(\theta )}.
\end{equation} 

The ultimate goal of any practical estimation procedure is at first to find the working regime maximizing the QFI, and then possibily finding a feasible configuration  to make the position FI equal to the QFI. In principle, the ultimate bound to precision may be achieved by measuring the SLD $\mathcal{L}_{\theta}$ which, however, may not correspond to a viable measurement scheme in the chosen implementation of DTQW.

\section{Understanding the role of the interference: the discrete line}
\label{sec:Interference}

To understand the role of interference we analyze the standard {DTQW} paradigm, where the position space is defined by an infinite one-dimensional discrete lattice, where $\mathscr{H}_p = \operatorname{span}\{\vert x \rangle_p \,\vert\, x \in \mathbb{N}, \; and \; x \leq \mathcal{N}-1\}$, and the coin space is bi-dimensional $\mathscr{H}_{c} = \{ \vert \pm 1 \rangle_{c} \}$. Under this assumption, the shift operator takes the form

\begin{align}
    \label{eq:conditional_shift_operator_1D_lattice}
    \mathcal{S} =  \sum_{x \in \mathbb{N}}  & \, \,
    \vert x + 1 \rangle_{p} {}_{p}\langle x \vert \otimes \vert 1 \rangle_{c}{}_{c}\langle 1 \vert
   \notag \\  &+
        \vert x - 1 \rangle_{p} {}_{p}\langle x \vert \otimes \vert - 1 \rangle_{c}{}_{c}\langle - 1 \vert,
\end{align}
setting at the boundary $\vert \mathcal{N} \rangle = \vert 1 \rangle$. For an initial probe localized in position space and with a general coin state the wave-function at $t=0$ reads

\begin{equation}
    \label{eq:localized_probe_state}
    \vert \psi (0) \rangle   = \vert x_{0} \rangle_{p} \otimes \Big( \alpha_{x_{0}} \vert -1 \rangle_c + \beta_{x_{0}} \vert +1 \rangle_c  \Big),
\end{equation}
where, under the normalization condition $\alpha_{x_{0}} \in \mathbb{R}$ and $\beta_{x_{0}} = e^{i\gamma_{x_{0}}} \sqrt{1- \abs{\alpha_{x_{0}}}^{2}}$, with $\gamma_{x_{0}} \in [0,2\pi]$. Interesting features arise upon exploring an enlarged set of initial states, which has the additional degree of freedom of a delocalized initial state of the walker, i.e., 
\begin{equation}
    \label{eq:general_probe_state}
    \vert \psi (0) \rangle   = \sum_{x} \vert x \rangle_{p} \otimes \left( \alpha_{x} \vert -1 \rangle_c + \beta_{x} \vert +1 \rangle_c  \right),
\end{equation}
where $\sum_{x} \left( \abs{\alpha_{x}}^2 + \abs{\beta_{x}}^2 \right) = 1 $.
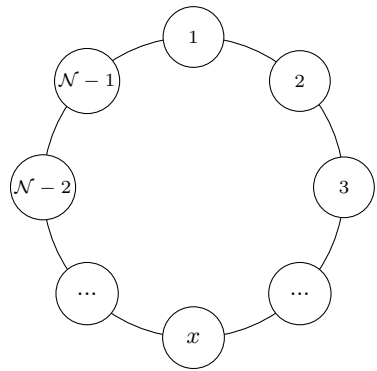
\begin{figure}[h!]
    \centering
    \begin{tikzpicture}
     \draw (0,0) circle (2cm);
     \node[fill=white, draw=black, circle, inner sep=6pt] at (45*1:2cm) {\scriptsize $2$};
     \node[fill=white, draw=black, circle, inner sep=6pt] at (45*2:2cm) {\scriptsize $1$};
     \node[fill=white, draw=black, circle, inner sep=1pt] at (45*3:2cm) {\scriptsize $\mathcal{N}-1$};
     \node[fill=white, draw=black, circle, inner sep=1pt] at (45*4:2cm) {\scriptsize $\mathcal{N}-2$};
     \node[fill=white, draw=black, circle, inner sep=6pt] at (45*5:2cm) {$...$};
     \node[fill=white, draw=black, circle, inner sep=6pt] at (45*6:2cm) {$x$};
     \node[fill=white, draw=black, circle, inner sep=6pt] at (45*7:2cm) {$...$};
     \node[fill=white, draw=black, circle, inner sep=6pt] at (45*8:2cm) {\scriptsize $3$};
    \end{tikzpicture}
    \caption{Standard topology of a discrete line with periodic boundary conditions for a {DTQW}. The labeling of the vertices is related to the definition of the adjacency matrix $\mathcal{A}^{b}$
    and shift operator $\mathcal{S}$,  see \ref{app:Topology}.}
    \label{fig:1D}
\end{figure}

\subsection{Encoding the parameter by z-rotations}
For a $z$-rotation encoding of the parameter, the coin operator 
is diagonal and may be written as
\begin{equation}
    \label{eq:Cz}
        \mathcal{C}_{z}(\theta)=\begin{pmatrix}
        e^{-i\theta/2} & 0 \\
        0 & e^{i\theta/2}
    \end{pmatrix}\,.
\end{equation}
The effect on the dynamics is a sequential addition of different phases 
to the $\vert \pm 1 \rangle$ component of the walker's wave-function. 
In detail, for an initially localized state of the form \eqref{eq:localized_probe_state} the wave-function at time $t$ is 
given by
\begin{align}
    \label{eq:psit_loc_Rz}
    \vert \psi(t) \rangle = & e^{-i \theta t / 2} \alpha_{x_{0}} \vert x_{0}-t\rangle_{p} \otimes \vert -1 \rangle_c \nonumber \\
    &+ e^{i \theta t / 2 } \beta_{x_{0}} \vert x_{0}+t\rangle_{p} \otimes \vert +1 \rangle_c.
\end{align}
Consequently, the (Q)FI at time step $t$ reads
\begin{equation}
    \label{eq:(q)fi_localized_Rz}
    \begin{cases}
        H^{x_{0}}_{z}(t) =   t^{2} \left[1 - \left(2\abs{\alpha_{x_{0}}}^2 -1 \right)^{2}  \right] \\ 
        F_{p}(t)=0,
    \end{cases}
\end{equation}
where the apex $x_{0}$ refer to the localization condition. The maximal (Q)FI attainable for a localized state is $H^{x_{0}}_{z}(t)=t^2$ and state that maximize the QFI is in the form

\begin{equation}
    \label{initial_loc_max_QFI_Rz}
    \vert \phi^{opt,z}_{x_{0}} \rangle_{c} = \frac{1}{\sqrt{2}} \left( \vert -1 \rangle_c + e^{i \gamma_{x_{0}}} \vert 1 \rangle_c \right).
\end{equation}
for each complex phase angle $\gamma_{x_{0}} \in \left[ 0,2\pi \right]$.

Relaxing the condition on the localization on the initial state $\vert \psi (0) \rangle$ and considering an initial state in the form of Eq. \eqref{eq:general_probe_state}, at time $t$ the wave-function reads

\begin{align}
    \label{eq:psit_Rz}
    \vert \psi(t) \rangle = & \sum_{x} e^{-i \theta t / 2} \alpha_{x} \vert x-t\rangle_p \otimes \vert -1 \rangle_c \nonumber \\
    &+ e^{i\theta t / 2} \beta_{x} \vert x+t\rangle_p \otimes \vert +1 \rangle_c.
\end{align}
This means that every reticular site with an associated initial probability different from zero originates an independent {DTQW} branch. The (Q)FI of the probe is then 
\begin{equation}
    \label{eq:(q)fi_Rz}
    \begin{cases}
        H_{z}(t) = t^{2} \left[1 - \left( \sum_{x} \abs{\alpha_{x}}^2 - \abs{\beta_{x}}^2 \right)^{2}  \right] \\
        F_{p}(t)=0.
    \end{cases}
\end{equation}
The maximal QFI attainable is again $H_{z}(t)=t^2$, associated to an optimal initial probe which fulfill the condition $ \left( \sum_{x} \abs{\alpha_{x}}^2 - \abs{\beta_{x}}^2 \right)^{2}=0 $, for any distribution in the position space. The only condition is then related to the coin's initial state of the probe. This means that the maximum value of the (Q)FI for a $z$-rotation encoding can be achieved both by a localized and a non localized initial state of the walker. Eq. \eqref{eq:(q)fi_Rz} can be understand intuitively from the structure of the coin matrix $\mathcal{C}_{z}$. Since it is diagonal, every branch of the {DTQW} does not interfere with the other ones and has an independent time evolution. Then, the localization condition in position space does not affect the global (Q)FI of the system. Position measurement is thus not suitable for inferring information about variables related to internal degrees of freedom of the walker. 

\subsection{Encoding the parameter by x- or y-rotations}

Unlike the $z$- rotation, $\mathcal{C}_{x}$ and $\mathcal{C}_{y}$ coin encodings are non diagonal in the standard basis, and share common features that differentiate the time evolution from that induced by $\mathcal{C}_{z}$ \cite{PhysRevA.109.022432}. 
For the $x$- encoding of the parameter we have
\begin{equation}
    \label{eq:Cx}
        \mathcal{C}_{x}(\theta)=\mathcal{V}\mathcal{C}_{z}(\theta)\mathcal{V}^{\dagger}=\begin{pmatrix}
        \cos{\theta/2} & i\sin{\theta/2} \\
        i\sin{\theta/2} & \cos{\theta/2}
    \end{pmatrix}\,,
\end{equation}
where $\mathcal{V}$ is a fixed rotation, and any state of the form
\begin{align}
    \label{eq:initial_loc_max_QFI_Rx}
    & \ket{\psi(0)} = \ket{x_{0}}_{p} \otimes \vert \phi^{opt,x}_{x_{0}} \rangle_{c}  = \ket{x_{0}}_{p} \otimes  \mathcal{V}\, \vert \phi^{opt,z}_{x_{0}} \rangle_{c} = \nonumber\\
    & = |x_0\rangle_p \otimes \left[ \alpha_{x_{0}}  \vert -1 \rangle_c \pm i \sqrt{1-\abs{\alpha_{x_{0}}}^{2}} \vert +1 \rangle_c \right].
\end{align}
represents an optimal, initially localized, preparation of the probe.
Similarly for a $y$-encoding, we have 
\begin{equation}
    \label{eq:Cy}
        \mathcal{C}_{y}(\theta)=\mathcal{W}\mathcal{C}_{x}(\theta)\mathcal{W}^{\dagger}=\begin{pmatrix}
        \cos{\theta/2} & -\sin{\theta/2} \\
        \sin{\theta/2} & \cos{\theta/2}
    \end{pmatrix}\,,
\end{equation}
and any real state of the form
\begin{align}
    \label{eq:initial_loc_max_QFI_Ry}
    & \ket{\psi(0)} = \ket{x_{0}}_{p} \otimes \vert \phi^{opt,y}_{x_{0}} \rangle_{c} = \ket{x_{0}}_{p} \otimes \mathcal{W}\, \vert \phi^{opt,z}_{x_{0}} \rangle_{c} = \nonumber \\
    & =  |x_0\rangle_p \otimes \left[ \alpha_{x_{0}}  \vert -1 \rangle_c \pm \sqrt{1-\abs{\alpha_{x_{0}}}^{2}} \vert +1 \rangle_c \right] \, ,
\end{align}
represents an optimal initial preparation of the initially localized probe and the $\mathcal{C}_{y}(\theta)$ coin.

Unlike for the $z$-encoding, the QFI for initially localized state does depend on the actual value of the parameter. We have 
\begin{equation}
\label{eq:limits_qfi}
    \lim_{\theta \to 0} H_{x_{0}}^{x,y}(\theta) \propto t, \quad \text{otherwise} \quad  H_{x_{0}}^{x,y} \propto t^2.
\end{equation}
such that the maximal QFI attainable is
\begin{align}
    \label{eq:max_qfi_rxy}
    H_{x_{0}}^{x,y} (t) = \frac{t^2}{2} + \frac{mod(t,2)}{2}\,, 
    \end{align}
corresponding to $\theta=\pi,3\pi$. The discontinuity of the QFI at $\theta=0$ is not surprising, since $\mathcal{C}_{x}(\theta)$ and $\mathcal{C}_{x}(\theta)$ both reduce to the identity for $\theta=0$. In this case the position and the coin degrees of freedom do not get entangled and the rank of the density matrix does not increase, as it happens for any $\theta\neq 0$ \cite{seveso20}.

The FI of the position is different from zero for $x$- and $y$-encodings. As matter of fact, it  is not possible to obtain a {\em universal} optimal state maximizing $F_{p}$ at any time and for any value of $\theta$. Nevertheless, the value of the FI is always a significant fraction of the QFI and we thus conclude that a position measurement represents a good strategy to infer information about $\theta$. The dependence of the position of the walker from the coin's degree of freedom arises from the non diagonal structure of the coin matrix, leading to an interference pattern in position space which depends both on the initial state of the probe, the coin matrix and the value of $\theta$.

In order to go beyond localized initial states, we consider a 
Gaussian-like state of the form 
\begin{equation}
    \label{eq:separable_initial_state}
    \vert \psi \left( 0 \right) \rangle = \sqrt{A} \sum_{x} e^{\frac{-(x-x_{0})^2}{2\sigma^2}} \vert x \rangle_p \otimes \vert \phi^{opt,x,y}_{x} \rangle_{c},
\end{equation}
centered in $x_{0}$ and with a variance $\sigma$, $A$ being a normalization constant. In the limit of an infinite sharp Gaussian, i.e. $\sigma \to 0$, the time evolution approaches the localized initial state, while in the limit of $\sigma \to \infty$ the amplitudes in position space approach a uniform distribution. 

\begin{figure}[h!]
    \includegraphics[width=0.95\columnwidth]{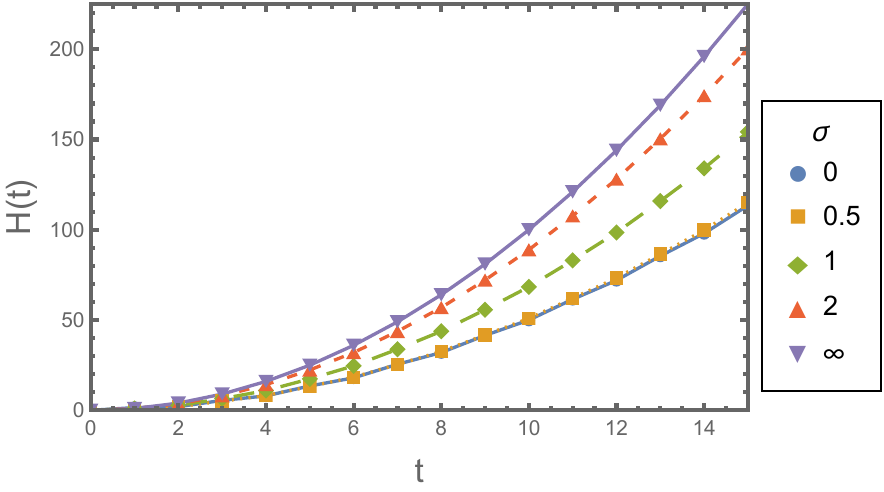}
    \caption{Time dependence of the QFI obtained for an initial Gaussian state in the form Eq. \eqref{eq:separable_initial_state}. The initial state in coin space is set as $\vert \phi^{y} \rangle = \vert -1 \rangle$, while the coin matrix is $C_{y}(\theta=\pi/2)$. Different colors denote the curves 
    obtained for different values of $\sigma$.}
    \label{fig:1_Gaussian_QFI}
\end{figure}

\begin{figure}[h!]
    \includegraphics[width=0.95\columnwidth]{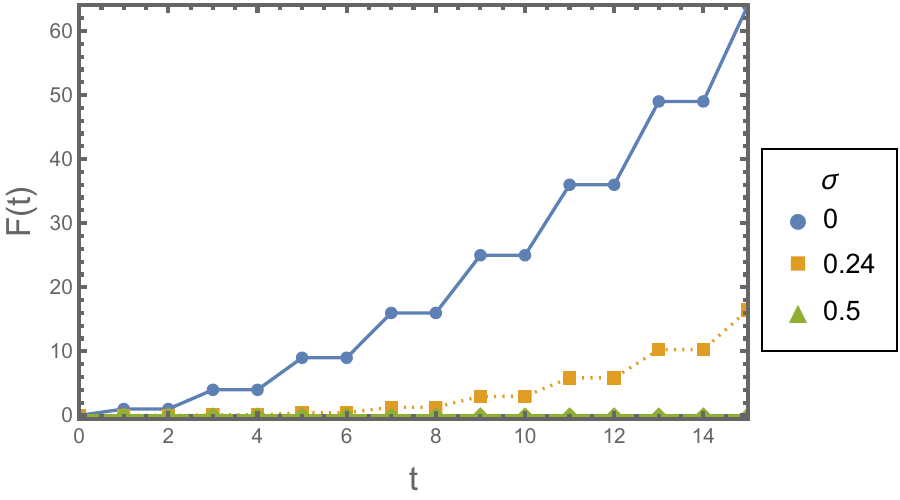}
    \caption{Time dependence of the FI obtained for an initial Gaussian state in the form Eq. \eqref{eq:separable_initial_state}. The initial state in coin space is set as $\vert \phi^{y} \rangle = \vert -1 \rangle$, while the coin matrix is $C_{y}(\theta=\pi/2)$. Different colors denote the curves 
    obtained for different values of $\sigma$.}
    \label{fig:2_Gaussian_FI}
\end{figure}

Upon considering this initial probe interesting features emerges. As the value of $\sigma$ increases the value of the Quantum Fisher information increases, approaching the maximal QFI attainable with a $z$- rotation (i.e. $H(t) =t^2$, see Fig.\ref{fig:1_Gaussian_QFI}). The Fisher information of a position measurement approaches the behavior of the FI of a $z$-rotation too (see Fig.\ref{fig:2_Gaussian_FI}), i.e.  it vanishes for increasing $\sigma$.  This may be understood by considering the time evolution of an initial probe in the form
\begin{equation}
    \label{eq:delocalized_initial_state}
    \vert \psi \left( 0 \right) \rangle = \frac{1}{\sqrt{\mathcal{N}}} \sum_{x}^{\mathcal{N}} \vert x \rangle_p \otimes \vert \phi^{opt,x,y}_{x} \rangle_{c},
\end{equation}
which, at time $t$, evolves to the following wave-function for $x$-rotations
\begin{align}
    \label{eq:delocalized_state_x_t}
    \vert \psi \left( t \right) \rangle = \frac{1}{\sqrt{\mathcal{N}}} \sum_{x}^{\mathcal{N}} \vert x \rangle_p \otimes  \frac{1}{\sqrt{2}} \biggl[ e^{-i\theta/2} \vert v_{1} \rangle_{c} + e^{i\theta/2 + \gamma } \vert v_{2} \rangle_{c} \biggl],
\end{align}
where $\mathcal{C}_{x} \vert v_{i} \rangle_{c} = v_{i} \vert v_{i} \rangle_{c}$ with $i=1,2$, and to
\begin{align}
    \label{eq:delocalized_state_y_t}
    \vert \psi \left( t \right) \rangle = \frac{1}{\sqrt{\mathcal{N}}} \sum_{x}^{\mathcal{N}} \vert x \rangle_p \otimes  \frac{1}{\sqrt{2}} \biggl[ e^{-i\theta/2} \vert w_{1} \rangle_{c} + i e^{i\theta/2 + \gamma} \vert w_{2} \rangle_{c} \biggl],
\end{align}
for $y$-rotations, where $\mathcal{C}_{y} \vert w_{i} \rangle_{c} = w_{i} \vert w_{i} \rangle_{c}$ with $i=1,2$. 


It follows that  in the limit of an uniform distribution, the DTQW with a $\mathcal{C}_{x}$ (or $\mathcal{C}_{y}$) coin approaches the metrological behavior of the $\mathcal{C}_{z}$ encoding (see the structure of Eq.\eqref{eq:psit_Rz} and of Eq.\eqref{eq:delocalized_state_x_t},\eqref{eq:delocalized_state_y_t}). 
Indeed, as the Gaussian distribution of the initial probe approaches the uniform distribution, the interference in position space disappear, and the position distribution becomes independent of the coin parameter(see Appendix \ref{app:relation_DTQW_QuDit}). As the Gaussian distribution approaches the localized state, we have the opposite behavior, and the position space distribution has the strongest dependence on the coin's degree of freedom. 

Overall, we conclude that for a one-dimensional lattice it is always possible to maximize the value of the QFI of a parameter encoded in a $D=2$ rotation. However, the corresponding position FI never approaches the QFI or even vanishes, and \eqref{eq:Quantum_Cramer_Rao} is strictly an inequality. 

\section{Topologically enhanced metrology with DTQW}
\label{sec:Topology}
In this Section, we consider modification of the graph topology in order to fully exploit the interference in the position degree of freedom, i.e. the coin-position entanglement. As we will see, by exploring the topology of the graph it is possible to enhance the metrological performance of the {DTQW}. This ideal topology have to maximize the QFI of the encoded parameter $\theta$ and to make the position FI equal to the QFI.

\subsection{Bi-dimensional Coin}
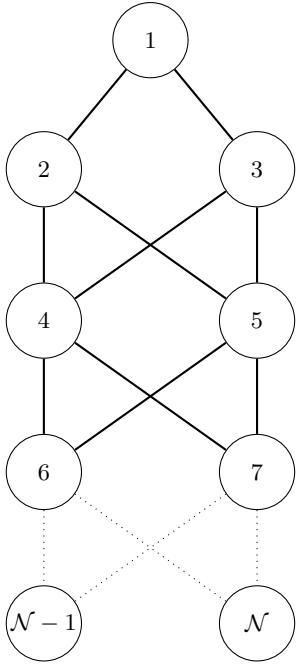
\begin{figure}[h!]
    \centering 
    \begin{tikzpicture}[scale = 0.5]
    \tikzset{main node/.style={circle,fill=white!20,draw,minimum size=1.0cm,inner sep=0pt}}
      \node[main node] (1) {$1$};
      \node[main node] (2) [below left = 1.0cm and 0.7cm of 1]  {$2$};
      \node[main node] (3) [below right = 1.0cm and 0.7cm of 1] {$3$};
      \node[main node] (4) [below = 1.0cm  of 2] {$4$};
      \node[main node] (5) [below = 1.0cm  of 3] {$5$};
      \node[main node] (6) [below = 1.0cm  of 4] {$6$};
      \node[main node] (7) [below = 1.0cm  of 5] {$7$};
      \node[main node] (8) [below = 1.0cm  of 6] {$\mathcal{N}-1$};
      \node[main node] (9) [below = 1.0cm  of 7] {$\mathcal{N}$};
      \path[draw,thick]
      (1) edge node {} (2)
      (1) edge node {} (3)
      (2) edge node {} (4)
      (2) edge node {} (5)
      (3) edge node {} (4)
      (3) edge node {} (5)
      (4) edge node {} (6)
      (4) edge node {} (7)
      (5) edge node {} (6)
      (5) edge node {} (7);
      \path[draw,dotted]
      (6) edge node {} (8)
      (6) edge node {} (9)
      (7) edge node {} (8)
      (7) edge node {} (9);
    \end{tikzpicture}
    \caption{Ideal topology for a bi-dimensional {DTQW}. The labeling of the vertices is related to the definition of the adjacency matrix $\mathcal{A}^{t.e.}$ and shift operator $\mathcal{S}$, see \ref{app:Topology}.}
    \label{fig:Ideal_Topology}
\end{figure}

Let us consider the graph of Fig.\ref{fig:Ideal_Topology}. The shift operator that describe the propagation of the {walker} reads as follows

\begin{align}
    \label{eq:conditional_shift_operator_ideal_topology}
    \mathcal{S}^{t.e}_{D=2} = & \sum_{x_{odd}} \vert x+1 \rangle_{p} {}_{p}\langle x \vert \otimes \vert -1 \rangle_{c} {}_{c}\langle -1 \vert + \nonumber \\
    &\vert x+2 \rangle_{p} {}_{p}\langle x \vert \otimes \vert +1 \rangle_{c} {}_{c}\langle +1 \vert \nonumber \\
    & + \sum_{x_{even}} \vert x+2 \rangle_{p} {}_{p}\langle x \vert \otimes \vert -1 \rangle_{c} {}_{c}\langle -1 \vert + \nonumber \\
    & \vert x+3 \rangle_{p} {}_{p}\langle x \vert \otimes \vert +1 \rangle_{c} {}_{c}\langle +1 \vert\,.
\end{align}
The  encoding of the parameter in the coin $\mathcal{C}(\theta)$ have to 
be non-diagonal, otherwise the position of the walker becomes 
independent of $\theta$ (i.e. $F_p=0$), and only the QFI is maximized. Mathematically, if we consider the $\mathcal{C}_{z}(\theta)$ coin, and we consider an initial wave-function localized in the $\ket{x_{0}}_{p}=\ket{1}_{p}$ node, the wave-function of the system at time $t$ reads

\begin{align}
    \label{eq:wavefunction_z_Ideal_topology}
        \vert \psi(t>0) \rangle = & e^{-i \theta t / 2} \alpha_{x_{0}} \vert 2 t \rangle_{p} \otimes \vert -1 \rangle_c \nonumber \\
        &+ e^{i\theta t / 2} \beta_{x_{0}} \vert 2t+1 \rangle_{p} \otimes \vert +1 \rangle_c,
\end{align}
with the (Q)FI completely analogous to Eq. \eqref{eq:(q)fi_localized_Rz}. On the other hand, using a non diagonal coin, as $\mathcal{C}_{x}$ or $\mathcal{C}_{y}$, and the shift operator defined in Eq. \eqref{eq:conditional_shift_operator_ideal_topology}, the wave-function at time $t$ reads
\begin{align}
    \label{eq:psit_optimal_topology}
    \vert \psi (t>0) \rangle = & {}_{c}\langle -1 \vert \mathcal{C}(\theta)_{x/y}^{t} \vert \phi \rangle_{c}  \vert 2t \rangle_{p} \otimes \vert -1 \rangle_{c} + \nonumber \\
    &{}_{c}\langle +1 \vert \mathcal{C}(\theta)_{x/y}^{t} \vert \phi \rangle_{c} \vert 2t+1 \rangle_{p} \otimes \vert +1 \rangle_{c} \nonumber \\
    = & {}_{c}\langle -1 \vert \mathcal{C}(t\theta)_{x/y} \vert \phi \rangle_{c}  \vert 2t \rangle_{p} \otimes \vert -1 \rangle_{c} + \nonumber \\
    &{}_{c}\langle +1 \vert \mathcal{C}(t\theta)_{x/y} \vert \phi \rangle_{c} \vert 2t+1 \rangle_{p} \otimes \vert +1 \rangle_{c}.
\end{align}
According to Eq, \eqref{eq:quantum_fisher_information_pure_states} the QFI for $x$-encoding is given by 

\begin{equation}
    \label{eq:qfi_x_ideal_topology}
    H^{x}_{x_{0}}(t)=t^{2}\left[1- 4\alpha_{x_{0}}^{2} (1-\alpha_{x_{0}}^{2})\cos^{2}\gamma_{x_{0}} \right],
\end{equation}
which is maximized for an initial probe in the form 

\begin{equation}
    \label{eq:initial_loc_max_QFI_Rx_ideal_topology}
    \vert \phi^{opt,x}_{x_{0}} \rangle = \alpha_{x_{0}}  \vert -1 \rangle_c \pm i  \sqrt{1-\abs{\alpha_{x_{0}}}^{2}} \vert +1 \rangle_c.
\end{equation}
Similarly, the QFI for $y$-encoding is 

\begin{equation}
    \label{eq:qfi_y_ideal_topology}
    H^{y}_{x_{0}}(t)=t^{2}\left[ 1- 4\alpha_{x_{0}}^{2} (1-\alpha_{x_{0}}^{2})\sin^{2}\gamma_{x_{0}} \right],
\end{equation}
which is maximized for an initial probe in the form 

\begin{equation}
    \label{eq:initial_loc_max_QFI_Ry_ideal_topology}
    \vert \phi^{opt,y}_{x_{0}} \rangle = \alpha_{x_{0}}  \vert -1 \rangle_c \pm \sqrt{1-\abs{\alpha_{x_{0}}}^{2}} \vert +1 \rangle_c.
\end{equation}
In both cases, considering an initial state in the form Eq.\eqref{eq:initial_loc_max_QFI_Rx_ideal_topology}-\eqref{eq:initial_loc_max_QFI_Ry_ideal_topology} the maximum value attainable saturates inequality Eq.\eqref{eq:Quantum_Cramer_Rao} and is given by
\begin{equation}
    \label{eq:qfi_max_ideal_topology}
    H_{x_{0}}^{x/y}(t)=F_{p}^{x/y}(t)=t^2.
\end{equation}

\subsection{D-dimensional Coin}
The optimal topology introduced for $D=2$ may be generalized to a discrete time quantum walk with a generic $D$-dimensional coin. The shift operator that describe the propagation of the {walker} on such topology reads

\begin{align}
    \label{eq:conditional_shift_operator_ideal_topology_D}
    \mathcal{S}^{t.e.}_{D} = \sum_{t=0} \sum_{r=1}^{D} \sum_{m=1}^{D} & \bigg[ \vert D(t+1-\delta_{r1})+m+1 \rangle_{p} {}_{p} \langle r+Dt \vert \nonumber \\
    & \otimes \vert m \rangle_{c} {}_{c} \langle m \vert \bigg]\,.
\end{align}
The generalized coin rotation matrices are

\begin{equation}
\label{eq:generators}
    \mathcal{C}^{(D)} _{\hat{n}} (\theta) = e^{-i \theta_{\hat{n}} \cdot \mathcal{T}^{(D)}_{\hat{n}}},
\end{equation}
which rotates the internal degree of freedom of the {walker} of an angle $\theta$ around the axis $\hat{n}=\hat{x},\hat{y},\hat{z}$. The matrices $\mathcal{T}^{(D)}_{\hat{n}}$ are the {generators} of the the rotation and are defined through the relations
\begin{equation}
\label{eq:lie}
\left[ \mathcal{T}^{(D)}_{a}, \mathcal{T}^{(D)}_{b} \right] = i \epsilon_{abc} \mathcal{T}^{(D)}_{c}
\end{equation}
where $ \epsilon_{abc} $ is the Levi-Civita symbol. 
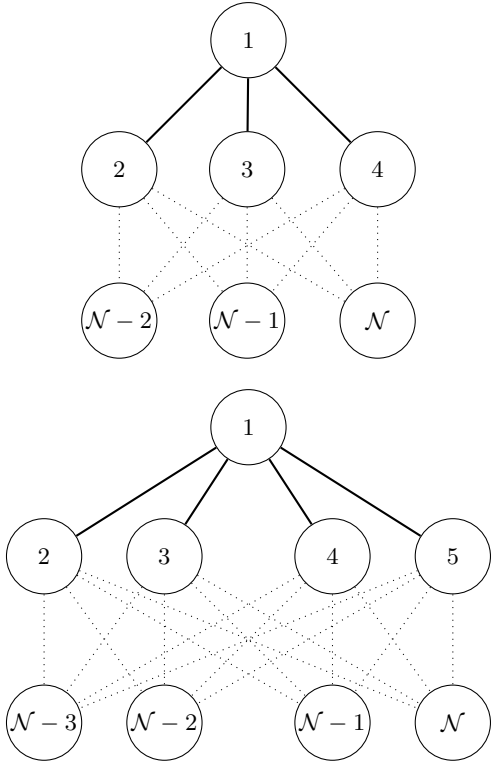
\begin{figure}[h!]
    \centering 
    \begin{tikzpicture}[scale = 0.5]
    \tikzset{main node/.style={circle,fill=white!20,draw,minimum size=1.0cm,inner sep=0pt}}
      \node[main node] (1) {$1$};
      \node[main node] (2) [below left = 1.0cm and 1.0cm of 1]  {$2$};
      \node[main node] (3) [below left = 1.0cm and -0.7cm of 1] {$3$};
      \node[main node] (4) [below right = 1.0cm and 1.0cm of 1] {$4$};
      \node[main node] (5) [below = 1.0cm of 2] {$\mathcal{N}-2$};
      \node[main node] (6) [below = 1.0cm of 3] {$\mathcal{N}-1$};
      \node[main node] (7) [below = 1.0cm of 4] {$\mathcal{N}$};
      \path[draw,thick]
      (1) edge node {} (2)
      (1) edge node {} (3)
      (1) edge node {} (4);
      \path[draw,dotted]
      (2) edge node {} (5)
      (2) edge node {} (6)
      (2) edge node {} (7)
      (3) edge node {} (5)
      (3) edge node {} (6)
      (3) edge node {} (7)
      (4) edge node {} (5)
      (4) edge node {} (6)
      (4) edge node {} (7);      
    \end{tikzpicture}

    \;

    \begin{tikzpicture}[scale = 0.5]
    \tikzset{main node/.style={circle,fill=white!20,draw,minimum size=1.0cm,inner sep=0pt}}
      \node[main node] (1) {$1$};
      \node[main node] (2) [below left = 1.0cm and 2.0cm of 1]  {$2$};
      \node[main node] (3) [below left = 1.0cm and 0.4cm of 1] {$3$};
      \node[main node] (4) [below right = 1.0cm and 0.4cm of 1] {$4$};
      \node[main node] (5) [below right = 1.0cm and 2.0cm of 1] {$5$};
      \node[main node] (6) [below = 1.2cm of 2] {$\mathcal{N}-3$};
      \node[main node] (7) [below = 1.2cm of 3] {$\mathcal{N}-2$};
      \node[main node] (8) [below = 1.2cm of 4] {$\mathcal{N}-1$};
      \node[main node] (9) [below = 1.2cm of 5] {$\mathcal{N}$};
      \path[draw,thick]
      (1) edge node {} (2)
      (1) edge node {} (3)
      (1) edge node {} (4)
      (1) edge node {} (5);
      \path[draw,dotted]
      (2) edge node {} (6)
      (2) edge node {} (7)
      (2) edge node {} (8)
      (2) edge node {} (9)
      (3) edge node {} (6)
      (3) edge node {} (7)
      (3) edge node {} (8)
      (3) edge node {} (9)
      (4) edge node {} (6)
      (4) edge node {} (7)
      (4) edge node {} (8)
      (4) edge node {} (9)
      (5) edge node {} (6)
      (5) edge node {} (7)
      (5) edge node {} (8)
      (5) edge node {} (9);
    \end{tikzpicture}
    \caption{Top (Bottom): Optimal topology for DTQW-based metrology 
    with a three- (four-) dimensional coin. The labeling of the vertices is related to the definition of the adjacency matrix $\mathcal{A}^{t.e.}$ and shift operator $\mathcal{S}$, see \ref{app:Topology}.}
    \label{fig:Ideal_Topology_D}
\end{figure}
In particular, the $z$-rotation in dimension $D$ is given by
\begin{align}
\label{eq:D_dimensional_z_coin}
    \mathcal{C}^{(D)} _{z} (\theta) = & \hbox{Diag}\{ e^{-im_{c}\theta},...,e^{im_{c}\theta} \} = \nonumber \\
    = &\hbox{Diag}\{ e^{-i(D-1)\theta/2},...,e^{i(D-1)\theta/2} \},
\end{align}
while the $x$- and $y$- rotation matrices are defined by basis change as follows
\begin{equation}
    \begin{cases}
        \label{eq:basis_change_x_D}
        \mathcal{C}^{(D)}_{x}= \mathcal{V}^{(D)} \mathcal{C}^{(D)}_{z}(\theta) \mathcal{V}^{(D),\dagger} \\ 
        \mathcal{C}^{(D)}_{y}= \mathcal{W}^{(D)} \mathcal{C}^{(D)}_{z}(\theta) \mathcal{W}^{(D),\dagger}
    \end{cases}
\end{equation}
The wave-function of the system, with the shift operator defined in Eq.\eqref{eq:conditional_shift_operator_ideal_topology_D} and a $D$ dimensional rotation coin, at time $t$ reads
\begin{align}
    \label{eq:psit_optimal_topology_D}
    \vert \psi (t) \rangle = \sum_{m=1}^{D} & \left[ {}_{c} \langle m \vert \mathcal{C}_{x/y/z}(t\theta) \vert \phi_{x_{0}} \rangle_{c} \right] \nonumber \\
    & \vert t(D-1)+m \rangle_{p} \otimes \vert m \rangle_{c}.
\end{align}
The (Q)FI is maximized for an initial probe in the form 
\begin{equation}
    \label{eq:initial_loc_max_QFI_Rx/y_ideal_topology_D}
    \vert \phi^{opt,x/y/z}_{x_{0}} \rangle_{c} = \frac{1}{\sqrt{2}} \left( \vert e_{min} \rangle_{c} + e^{i\gamma_{x_{0}}}\vert e_{max} \rangle_{c} \right),
\end{equation}
where $\vert e_{min} \rangle_{c} $ ($\vert e_{max} \rangle_{c} $) represent the eigenvector associated to the minimum (maximum) eigenvalue of the $\mathcal{C}^{(D)}_{x/y/z}$ matrix. For all the three possible, $x-$ $y-$ and $z-$ rotation encodings, the maximum value of the QFI attainable is the same, and reads
\begin{equation}
    \label{eq:qfi_max_ideal_topology_D}
    H_{x_{0}}^{x/y/z}(t)=4 \abs{\partial_{\theta} \psi(t)}^{2}=(D-1)^2 t^2,
\end{equation}
i.e. when $ \langle \partial_{\theta} \psi(t) \vert \psi(t) \rangle $ does not have any complex component. The FI derived from Eq.\eqref{eq:psit_optimal_topology_D} is 

\begin{align}
    \label{eq:fisher_information_te_DTQW}
    F_{p}(t) = & \sum_{x \in \mathcal{G}} \frac{\left[ \partial_{\theta} \abs{\psi_{x}(t)}^{2} ) \right]^{2}}{\abs{\psi_{x}(t)}^{2}} = \nonumber \\
    = & \sum_{x \in \mathcal{G}} \frac{\left[ \langle \partial_{\theta} \psi_{x}(t) \vert \psi_{x}(t) \rangle + \langle \psi_{x}(t) \vert \partial_{\theta} \psi_{x}(t) \rangle ) \right]^{2}}{\abs{\psi_{x}(t)}^{2}} = \nonumber \\
    = & \sum_{x \in \mathcal{G}} \frac{\left[ 2 \hbox{Re} \langle \partial_{\theta} \psi_{x}(t) \vert \psi_{x}(t) \rangle \right]^{2}}{\abs{\psi_{x}(t)}^{2}},
\end{align}
which vanishes for any $D$ in the $z$- case. For non-diagonal coins (i.e. the $x$- and $y$- case), $F_{p}(t)$ is maximized when $ \langle \partial_{\theta} \psi(t) \vert \psi(t) \rangle $ is real and is given by
\begin{equation}
    \label{eq:max_fisher_information_te_DTQW}
    F_{p}(t) = 4 \abs{\partial_{\theta} \psi(t)}^{2},
\end{equation}
saturating the inequality Eq.\eqref{eq:Quantum_Cramer_Rao}
\begin{equation}
    \label{eq:(q)fi_max_ideal_topology_D}
    H_{x_{0}}^{x/y/z}(t)=F^{x/y}_{p}(t)=(D-1)^2 t^2.
\end{equation}

\section{Summary and conclusions}
\label{sec:Summary}
We have investigate the use of DTQWs as quantum probes in estimation problems where the parameter of interest is encoded in the internal degree of freedom of the walker, and derived an optimal topology of the walker's space to maximize the QFI and make the position FI equal to the QFI. 

To this aim, we have first analyzed the metrological performance of {DTQW}s 
on the line. We have considered different encodings, seeking for the optimal initial preparation of the walker, and found that the QFI may be maximized by both initially localized and delocalized walkers. On the other hand, we have found that the position FI is always strictly smaller than the QFI, or even vanishes in several configurations.

From the understanding of the effect of the interference on the (Q)FI of the system, we have then proposed a graph topology to enhance the performance of a general {DTQW}, proving that the suggested topology makes it is possible to achieve the ultimate precision attainable for any metrological problem involving {DTQW}. Our results confirm the role of coin dimension in DTQW quantum metrology \cite{PhysRevA.109.022432,Candeloro2024dimension} and may find application in DTQW implementations where some form of control of the topology is possible, e.g., wave-guides \cite{zhu2020photonic,di2023ultra} or circuital implementations \cite{acasiete2020implementation,wing2023circuit}.

\section*{Acknowledgements}
This work has been done under the auspices of GNFM-INdAM and has been partially supported by MUR through the project PRIN22-2022T25TR3-RISQUE and by MUR and EU through the project PRIN22-PNRR-P202222WBL-QWEST.
The authors are grateful to Giovanni Ragazzi for several fruitful discussions about the dynamics of DTQWs.

\appendix

\section{The adjacency matrices and the conditional shift operators}
\label{app:Topology}
To define the time evolution operator $\mathcal{U}$ in a {DTQW} paradigm, it is also necessary to define the adjacency matrix $\mathcal{A}$ of the graph in which the walker propagates in time. The adjacency matrix mathematically define the topology of the graph and the connections among the lattice sites in the discrete position space. The most general form of $\mathcal{A}$ reads

\begin{equation}
    \label{eq:adjacency_matrix}
    \mathcal{A} = \sum_{x} \sum_{n_{x}=1}^{D_{x}} \vert x + f(n_{x}) \rangle_{p} {}_{p}\langle x \vert,
\end{equation}
where $n_{x}$ is the index that define the nearest neighbors ($NN$) sites of $\vert x \rangle_{p}$, $D_{x}$ indicates the total number of $NN$ and $f(n_{x}$ is the function that define the possible transition from the vertex $x$. Then, to define a time independent shift operator for a regular graph ( i.e. $D(x)=D, \; \forall \; \vert x \rangle_{p}$), each term of the sum over the $NN$ is associated to a coin degree of freedom labelled by $m$, as

\begin{equation}
    \label{eq:adjacency_shift}
    \mathcal{S}=\sum_{x} \sum_{m} \vert x + f(m) \rangle_{p} {}_{p}\langle x \vert  \otimes \vert m \rangle_{c} {}_{c}\langle m \vert,
\end{equation}
where there is a correspondence between the coin degree of freedom and the nearest neighbour vertices labeling, 

\begin{equation}
    \label{eq:correspondence}
    n \leftrightarrow m.
\end{equation}
In Sec.\ref{sec:Interference} we focus on the discrete one dimensional lattice with periodic boundary conditions, which has an adjacency matrix in the form

\begin{equation}
    \label{eq:adjacency_matrix_line}
    \mathcal{A}^{l} = \sum_{x} \vert x \pm 1 \rangle_{p} {}_{p}\langle x \vert.
\end{equation}
For the topology proposed in Sec.\ref{sec:Topology} the adjacency matrix of a {topologically enhanced (t.e.) DTQW} with a bi-dimensional coin the graph structure reads

\begin{align}
    \label{eq:adjacency_matrix_bi_te}
    \mathcal{A}^{t.e.}_{D=2} = & \sum_{x_{odd}} \vert x+1 \rangle_{p} {}_{p}\langle x \vert + \vert x+2 \rangle_{p} {}_{p}\langle x \vert \nonumber \\
    & + \sum_{x_{even}} \vert x+2 \rangle_{p} {}_{p}\langle x \vert + \vert x+3 \rangle_{p} {}_{p}\langle x \vert.
\end{align}

The general adjacency matrix for a {topologically enhanced DTQW} with a $D$ dimensional coin is defined as

\begin{align}
    \label{eq:adjacency_matrix_d_te}
    \mathcal{A}^{t.e.}_{D} = & \sum_{t=0} \sum_{r=1}^{D} \sum_{m=1}^{D} \vert D(t+1-\delta_{m1})+m+1 \rangle \langle r+Dt \vert
\end{align}

\section{The coin matrices and the time evolution operators}
\label{app:Basis_Change}

Throughout all the work we considered as coin matrices the $x$, $y$ and $z$ rotations. Even if those matrices are related through a basis change, they generate different time evolution operators. Starting from the coin $\mathcal{C}_{z}$ the other coin matrices can be obtained as

\begin{equation}
    \label{eq:basis_change_x}
    \mathcal{C}_{x}= \mathcal{V} \mathcal{C}_{z} \mathcal{V}^{\dagger}
\end{equation}
and

\begin{equation}
    \label{eq:basis_change_y}
    \mathcal{C}_{y}= \mathcal{W} \mathcal{C}_{z} \mathcal{W}^{\dagger},
\end{equation}
with $\mathcal{V}$ and $\mathcal{W}$ that are defined through the relation between Pauli matrices as
\begin{equation}
    \label{eq:Pauli_Matrices}
        \sigma_{x}=\mathcal{V} \sigma_{z} \mathcal{V}^{\dagger} \; \; ; \; \; \sigma_{y}=\mathcal{W} \sigma_{y} \mathcal{W}^{\dagger},
\end{equation}
since $\mathcal{C}_{i}=e^{-i\sigma_{i} \cdot \theta/2}$. The global time evolution operator for a {DTQW} does not depends only on the coin matrix, but also on the shift operator $\mathcal{S}$. This component depends both on the structure of the position space, as analyzed in Appendix \ref{app:Topology}, but also on the eigenvectors of a coin matrix. The most natural choice is the eigenvector of the $z$ spin operator $S_{z}$ (see Eq.\eqref{eq:conditional_shift_operator}). Nonetheless is is also possible to write a conditional shift evolution operator that depends on other eigenstates as

\begin{equation}
    \label{eq:conditional_shift_operator_x}
    \mathcal{S}_{x} = \sum_{x} \sum_{v} \vert f\left( x,t,v \right) \rangle_{x} \;_{x}\langle x \vert \otimes \vert v \rangle_{c} \;_{c}\langle v \vert,
\end{equation}
where $\mathcal{C}_{x} \vert v \rangle_{c} = v \vert v \rangle_{c}$ or

\begin{equation}
    \label{eq:conditional_shift_operator_{y}}
    \mathcal{S}_{y} = \sum_{x} \sum_{w} \vert f\left( x,t,w \right) \rangle_{x} \;_{x}\langle x \vert \otimes \vert w \rangle_{c} \;_{c}\langle w \vert,
\end{equation}
with $\mathcal{C}_{y} \vert w \rangle_{c} = w \vert w \rangle_{c}$. Starting from the time evolution operator with the $z$-rotation coin and the shift operator that depends on the eigenvalues of $S_{z}$

\begin{equation}
    \label{eq:time_evolution_z_coin}
    \mathcal{U}_{zz}= \mathcal{S}_{z} \left( \mathbb{1}_{p} \otimes \mathcal{C}_{z} \right),
\end{equation}
through a basis change we can obtain the following operators

\begin{align}
    \label{eq:time_evolution_x_shift_x_coin}
    \mathcal{U}_{x,x}= & \left( \mathbb{1}_{p} \otimes \mathcal{V} \right) \mathcal{S}_{z} \left( \mathbb{1}_{p} \otimes \mathcal{C}_{z} \right) \left( \mathbb{1}_{p} \otimes \mathcal{V}^{\dagger} \right) = \nonumber \\
    = & \mathcal{S}_{x} \left( \mathbb{1}_{p} \otimes \mathcal{C}_{x} \right),
\end{align}
and analogously

\begin{align}
    \label{eq:time_evolution_y_shift_y_coin}
    \mathcal{U}_{y,y}= & \left( \mathbb{1}_{p} \otimes \mathcal{W} \right) \mathcal{S}_{z} \left( \mathbb{1}_{p} \otimes \mathcal{C}_{z} \right) \left( \mathbb{1}_{p} \otimes \mathcal{W}^{\dagger} \right) = \nonumber \\
    = & \mathcal{S}_{y} \left( \mathbb{1}_{p} \otimes \mathcal{C}_{y} \right).
\end{align}
On the other hand if we start from a time evolution operator with the $x$- or $y$-rotation coin 

\begin{equation}
    \label{eq:time_evolution_x/y_coin}
    \mathcal{U}_{z,x/y}= \mathcal{S} \left( \mathbb{1}_{p} \otimes \mathcal{C}_{x/y} \right),
\end{equation}
through a basis change relation we can obtain the operator

\begin{align}
    \label{eq:time_evolution_z_shift_x_coin}
    \mathcal{U}_{x,z}= & \left( \mathbb{1}_{p} \otimes \mathcal{V} \right) \mathcal{S}_{z} \left( \mathbb{1}_{p} \otimes \mathcal{C}_{x} \right) \left( \mathbb{1}_{p} \otimes \mathcal{V}^{\dagger} \right) = \nonumber \\
    = & \mathcal{S}_{x} \left( \mathbb{1}_{p} \otimes \mathcal{C}_{z} \right)
\end{align}
and consequently

\begin{align}
    \label{eq:time_evolution_z_shift_y_coin}
    \mathcal{U}_{y,z}= & \left( \mathbb{1}_{p} \otimes \mathcal{W} \right) \mathcal{S}_{z} \left( \mathbb{1}_{p} \otimes \mathcal{C}_{y} \right) \left( \mathbb{1}_{p} \otimes \mathcal{W}^{\dagger} \right) = \nonumber \\
    = & \mathcal{S}_{y} \left( \mathbb{1}_{p} \otimes \mathcal{C}_{z} \right).
\end{align}
With the conditional shift operator fixed as $\mathcal{S}_{z}$, different coins $\mathcal{C}_{x/y/z}$ define different time evolution operators, not related through basis change transformations. Then, throughout the whole paper we set $\mathcal{S} = \mathcal{S}_{z}$ and define time evolution operators through the coin matrix

\begin{equation}
    \label{eq:final_time_evolution}
    \mathcal{U} = \mathcal{S}_{z} \left( \mathbb{1} \otimes \mathcal{C}_{x/y/z} \right)= \mathcal{S} \left( \mathbb{1} \otimes \mathcal{C}_{x/y/z} \right).
\end{equation}

\section{Relation between DTQW and qubit metrology}
\label{app:relation_DTQW_QuDit}

The behavior of the DTQW with a diagonal coin $\mathcal{C}_{z}$ and in the limit of a uniform distribution for the $\mathcal{C}_{x/y}$ rotations suggest that in both cases there is a relation among the behavior of the QFI of a DTQW and the QFI of a qubit (or qudit) \cite{metq0,metq1,metq2}.

{(i) The $z$- rotation case --} The time evolution of the walker with a diagonal coin can be written as 

\begin{align}
    \label{eq:time_evolution_diag_coin}
    \vert \psi (t) \rangle = & \mathcal{U}^{t} \vert \psi (0) \rangle = (\mathcal{S}(\mathbb{1} \otimes \mathcal{C}_{z}))^{t} \vert \psi (0) \rangle = \nonumber \\
    = & \mathcal{S}^{t} \left( \mathbb{1} \otimes \mathcal{C}_{z} \right)^{t} \vert \psi (0) \rangle,
\end{align}
then, proceeding to the calculations of the square modulus of the wave-function or to the scalar product between the $\vert \psi(t) \rangle$ and its derivative the shift operator does not play any role, as

\begin{align}
    \label{eq:scalar_product_diag_coin}
    & \langle \psi(t) \vert \psi (t) \rangle = \langle \psi(0) \vert (\mathcal{U}^{t})^{\dagger} \mathcal{U}^{t} \vert \psi (0) \rangle = \nonumber \\
    & = \langle \psi(0) \vert ((\mathcal{S}(\mathbb{1} \otimes \mathcal{C}_{z}))^{t})^{\dagger} (\mathcal{S}(\mathbb{1} \otimes \mathcal{C}_{z}))^{t} \vert \psi (0) \rangle = \nonumber \\
    & = \langle \psi(0) \vert (\mathbb{1} \otimes \mathcal{C}_{z})^{t})^{\dagger} (\mathbb{1} \otimes \mathcal{C}_{z})^{t} \vert \psi (0) \rangle
\end{align}
under the condition $\mathcal{S}^{\dagger}\mathcal{S}=\mathbb{1}$. Analogously

\begin{align}
    \label{eq:scalar_product_diag_coin_der}
    & \langle \psi(t) \vert \partial \psi (t) \rangle = \nonumber \\
    & = \langle \psi(0) \vert (\mathbb{1} \otimes \mathcal{C}_{z})^{t})^{\dagger} \partial(\mathbb{1} \otimes \mathcal{C}_{z})^{t} \vert \psi (0) \rangle.
\end{align}
Then, if the conditional shift and the coin part does commute, $\left[\mathcal{S},\mathbb{1} \otimes \mathcal{C} \right]=0$ the time evolution of the system is not influenced by the conditional shift operator, and depends only on the initial state of the system and on the operator $\mathbb{1} \otimes \mathcal{C}$. Upon this condition the QFI of the system is the QFI of a superposition of different qubit states (each one associated to a discrete position lattice site) under the time evolution of $\mathbb{1} \otimes \mathcal{C}$ (in the paper this condition is obtained using $\mathcal{C}_{z}$). The FI of a position measurement is identically null since the position distribution of the walker does not affect any physical quantity.

{(ii) The $x$- and $y$- rotation case --} For the $x$- and $y$- rotations the conditional shift part of the time evolution operator do not commute with the coin part. Nonetheless it is possible to recall the same situation as in the diagonal coin case with a condition on the initial state as

\begin{align}
    \label{eq:non_diagonal_coin_condition}
    \vert \psi(t) \rangle = & (\mathcal{S}(\mathbb{1} \otimes \mathcal{C}_{x/y}))^{t} \vert \psi(0) \rangle = \nonumber \\
     = &  (\mathcal{S}(\mathbb{1} \otimes \mathcal{C}_{x/y}))^{t} \left[ \mathcal{F}\vert x \rangle_{p} \otimes \vert \phi \rangle_{c} \right]= \nonumber \\
     = & \mathcal{F}\vert x \rangle_{p} \otimes \mathcal{C}_{x/y}^{t} \vert \phi \rangle_{c},
\end{align}
where the conditional shift operator evolve the state $\mathcal{F}\vert x \rangle_{p}$ into itself at each time-step. Under this consideration, and under the condition $\mathcal{F}^{\dagger}\mathcal{F}=\mathbb{1}$, the scalar product of the wave-function with itself or with its derivative assume the same form as Eq. \eqref{eq:scalar_product_diag_coin}-\eqref{eq:scalar_product_diag_coin_der}. For the topology of a line this condition is fulfilled by the equal superposition of each vertex, as derived in Sec.\ref{sec:Interference}.

\bibliography{Otfqpwd-tqw.bib}

\end{document}